\documentclass[twocolumn,prl,twocolumn,groupedaddress]{revtex4}
\usepackage{dcolumn}
\usepackage{bm}

\usepackage{epsfig,citesort} 
\usepackage{amsmath} 
\usepackage{amssymb}  
\usepackage{natbib}
\usepackage{wrapfig}

\newcommand{\utwi}[1]{\mbox{\boldmath $ #1$}}

\newcommand{\be}{{\utwi{e}}}

\newcommand{\bn}{{\utwi{n}}}

\newcommand{\bp}{{\utwi{p}}}

\newcommand{\bv}{{\utwi{v}}}

\newcommand{\bH}{{\utwi{H}}}
\newcommand{\bI}{{\utwi{I}}}

\newcommand{\bM}{{\utwi{M}}}

\newcommand{\bR}{{\utwi{R}}}

\newcommand{\bV}{{\utwi{V}}}

\begin{document}
\title{
Protein folding dynamics via quantification of kinematic
energy landscape
}

\author{S\"{e}ma Kachalo}
\author{Hsiao-Mei Lu}
\author{Jie Liang$^*$} 
\affiliation{Department of Bioengineering, MC-063,
University of Illinois at Chicago, Chicago, IL, 60305, U.S.A.
}

\date{
\today, 
Accepted by {\it Phys. Rev. Lett.}
}

\begin{abstract}
We study folding dynamics of protein-like sequences on square lattice
using physical move set that exhausts all possible conformational
changes.  By analytically solving the master equation, we follow the
time-dependent probabilities of occupancy of all 802,075 conformations
of 16-mers over 7-orders of time span.  We find that (i) folding
rates of these protein-like sequences of same length can differ by
4-orders of magnitude, (ii) folding rates of sequences of the same
conformation can differ by a factor of 190, and (iii) parameters of
the native structures, designability, and thermodynamic properties are
weak predictors of the folding rates, rather, basin analysis of the
kinematic energy landscape defined by the moves can provide
excellent account of the observed folding rates.
\end{abstract}

\maketitle
The dynamics of protein folding has been studied extensively
\cite{PlaxcoBaker-JMB,rateModel}.
A remarkable observation is that protein folding rates are
well correlated with their native structural properties
\cite{PlaxcoBaker-JMB}. A native centric view
postulates that protein folding rates are largely determined
by the topology of its native structure
\cite{GillespiePlaxco}. Theoretical models using G\={o} potential
where only native contacts contribute energetically are
very successful in reproducing observed folding rates
\cite{rateModel,WeiklDill}.

However, the extent to which native structure determines folding rate
remains unclear.  By the native-centric view, different sequences for
the same protein structural fold would all have very similar folding
rates, as they share essentially the same native structure topology.
However, this is not consistent with experimental results.  As
the folding rates of simple single-domain proteins differ by 6 orders
of magnitude \cite{GillespiePlaxco},  folding rates may be very
heterogeneous.  
A recent experimental study showed that a designed artificial protein
with no homologous sequence in nature that adopts the same structure
as a natural protein can fold 4,000 times faster
\cite{Scalley-Kim-Baker-04}.
A distinct possibility is that the empirical correlation between
properties of native protein structures and folding rates 
may arise from inadequate sampling in the sequence space due to
accumulated biased natural selection and limited genetic drift, rather
than from intrinsic physical properties of proteins.

In this letter, we use two-dimensional hydrophobic and polar (HP)
lattice model \cite{ChanDill94} to study the relationship of folding
rates, native structure topology, thermodynamic properties, and
effects of sequence variation.  We model the physical movement of
protein chains.  Real protein cannot immediately jump from one
conformation to another arbitrary conformation.  Two conformations of
the same energy may be well separated kinetically.  We regard
protein movement as a sequence of successive conformational changes,
each represented by a physically realizable primitive move.  The
physical move set we developed exhausts all possible conformational
changes for a structure.  We use master equation to study the folding
dynamics of foldable sequences of length 16.  While master equation
provides an exact solution \cite{ChanDill94,Cieplak98-PRL}, in the
past it was necessary to cluster conformations of larger systems into
macrostates to reduce the size of the transition matrix
\cite{OzkanBaharDill-NSB}, therefore making the use of physical moves
infeasible.  
Here we
directly solve the eigenvalue problem of the 802,075 $\times$ 802,075
transition matrix and develop a method to monitor the time-dependent
probability of occupancy of all conformations simultaneously from the
first kinetic move until reaching half-equilibrium concentration over
7-orders of time scale.

Our results show that the properties of native structure, designability,
and thermodynamic properties are inadequate to explain protein folding
dynamics in our model systems.  We found that protein-like sequences
can fold into the same native structure with folding rates differ as
much as 190 times and sequences of the same length and energy
gap can differ by 4-orders of magnitude in folding rate.  Instead of
thermodynamic properties, we show that properties of the move-connected energy
landscape defined by the connection graph of physical moves can
provide excellent account for observed folding rates.

\begin{center}
\begin{figure}[h]
\includegraphics[width=2.5in]{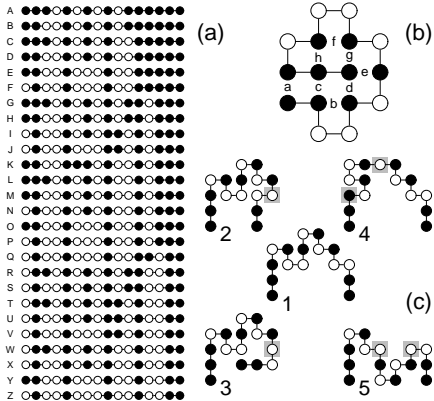}
\caption{\sf Protein-like sequences and the set of primitive moves.
The largest protein family contains (a) 26 sequences that all fold
into (b) the same structure. Here filled circles are H residues. (c) The
move set includes: among (1, 2, and 3), single point moves rotate
around a single point; between (1 and 4), generalized corner moves 
reflect around a diagonal axis connecting any two residues; between (1
and 5), generalized crankshaft moves  reflect around a
horizontal or vertical axis. Points of rotation are on gray
background. For a given conformation, we exhaustively search all
possible position for point moves, all possible pairs of 
positions for possible generalized corner moves and generalized
crankshaft moves.
\label{fig:seqMove}}
\end{figure}
\end{center}

{\bf Model.} We use the following energy model for different types of
nonbonded HP contacts: $E_{HH}=1$, $E_{HP}=0$, and $E_{PP}=0$.  By
evaluating the energy level of all $2^{16}$ sequences of 16-mers on
all enumerated $|\Omega| = 802,075$ conformations, we have identified
26 sequences that all fold into the same ground state conformation
(Fig.~\ref{fig:seqMove}).  This set of sequences forms the largest
protein family, where each sequence adopts the same conformation, and
all are connected by (a series of) point mutations. Altogether, there
are 1,539 foldable sequences with unique ground state
conformations. There are 456 conformations that are the unique ground
state for 1 or more foldable sequences.

We develop a set of physically possible primitive
moves (Fig.~\ref{fig:seqMove}c). They are generalizations of corner
move, crankshaft move, and pivot move. We exhaust all possible
occurrence of such moves for every conformation.  We verified that
this move set is ergodic, {\it i.e.}, all conformations are connected
to each other by a series of primitive moves.  With this move set, the
simple energy scheme of the HP model leads to a complex energy
landscapes, with numerous local minima for a foldable sequence.

\vspace*{-9mm}

We use Metropolis-type of dynamics to assign the transition rate $r_{ij}$
from conformation $i$ to a neighbor conformation $j$ connected by a
move: $ r_{ij}= 1$ if $E(j)\le E(i)$; $r_{ij}= e^{-[E(j)-E(i)]/T}$
if $E(j)> E(i)$; and $r_{ij} =  - \sum_{i \ne k} r_{ik}$, if $j =
i$. For non-neighbors, $r_{ij} = 0$.  
We assume the
effects of viscosity and friction are negligible.

We follow \cite{Cieplak98-PRL,OzkanBaharDill-NSB} and use a master
equation to study protein folding dynamics.  Let $p_{i}(t)$ be the
probability that the HP molecule takes the $i$-th conformation at the
time $t$, then $d p_{i}(t)/dt= \sum_{i\neq j}[r_{ji}p_{j}(t)-
r_{ij}p_{i}(t)]$.  Written in vector form, we have: $ d \bp (t)/dt =
\bR \bp(t), $ where $\bR$ is the rate matrix whose entries are defined
by the above expression.  We choose temperature $T = 0.2$ in unit of
$\Delta E_{HH}/k_B$, which is below the folding temperature $T_f$ when
50\% of molecules take the native conformation.  $T_f$ varies from
$\sim 0.2$ to $\sim 0.5$ for different sequences.

A general solution of the master equation can be written as $\bp(t) =
\sum_i C_i \bn_i e^{-\lambda_i t}$ with $C_i = \bv_i^T \bp(0)$, where
$\lambda_i$ is the $i$-the eigenvalue of the rate matrix $\bR$,
$\bn_i$ the corresponding right eigenvector, $\bv_i$ the left
eigenvector, and $\bp(0)$ the initial vector of distribution of
conformations.  In this study, we use the high temperature condition
and assign $\bp(0) = {\bold 1}/|\Omega|$.  Two eigenvalues are of
particular interest: $\lambda_0 = 0$ corresponds to the equilibrium
Boltzmann distribution, and the smallest none-zero eigenvalue
$\lambda_1$ determines the slowest mode of relaxation. Following
\cite{OzkanBaharDill-NSB}, we take $\lambda_1$ as the folding rate
$k_f$ of the protein.  Although the full computation of all
eigenvalues and eigenvectors for a $802,075 \times 802,075$ matrix
$\bM$ is infeasible, $\lambda_1$ and the corresponding eigenvectors
$\bn_1$ and $\bv_1$ can be computed by an Arnoldi method.

\begin{center}
\begin{figure}[h]
\includegraphics[width=2.in]{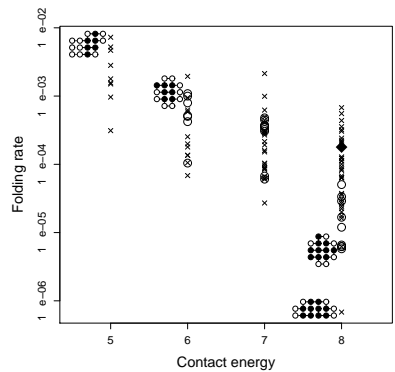}
\caption{\sf The correlation of $\log k_f$ and ground state contact
energy.  A circle represents one of the 26 sequences shown in
Fig.~\ref{fig:seqMove}, a cross represents one of the 79 singleton
sequences, and the diamond represents the G\={o} model. Native conformations
for a few sequences are also shown. }
\label{fig:stability}
\end{figure}
\end{center}
\vspace*{-9mm}

{\bf Thermodynamics and folding rates.} 
Several thermodynamic properties have been proposed to be determinants of
protein folding rates.  We found that protein stability as measured by
the total contact energy are correlated with $\log k_f$ ($R^2=0.71$),
{\it i.e.}, more stable proteins fold slower in general
(Fig.~\ref{fig:stability}).  
Because stable proteins have lower ground state energy, some
local minima will also have relatively deep energy traps.  As a
result, more stable proteins will have slower folding rates because
they can be trapped in such local minima.
However, the folding rates of sequences of
the same ground state energy can still differ as much as $10^4$.  The
heterogeneity of folding rate was already noted in an earlier study
using macrostate approximation
\cite{ChanDill94}. Here we found that even sequences that fold
into the same conformation shown in Fig.~\ref{fig:seqMove}a demonstrate
a wide range of rates, from $1.1\times10^{-3}$ to $5.8\times10^{-6}$,
which is much larger than the difference between the average folding
rates for sequences of different native state energies.  Protein
stability therefore provide some but not the main explanation of the
heterogeneity of folding rates.

Energy gap between ground state and excited state was thought to be
the necessary and sufficient determinant of folding rate
\cite{Sali94_Nature}.  For all 1,456 protein-like sequences of $N=16$,
the energy gap between the lowest state and the next state is $\Delta
E = 1$.  The diversity in folding rate $k_f$ shown in
Fig.~\ref{fig:stability} clearly indicates that energy gap is not a
determining factor for the folding rate.  
The correlation $R^2$
between $\log k_f$ and energy gap normalized by standard deviation is
0.01.

\begin{center}
\begin{figure}[h]
\includegraphics[width=3in]{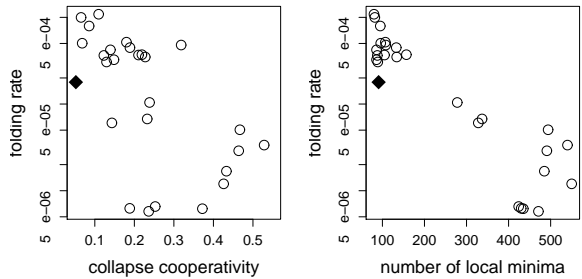}
\caption{\sf Examples of the correlation of folding rate $k_f$ with
thermodynamic properties and kinetic landscape properties.  (a)
$k_f$ and collapse cooperativity $\sigma$ have weak correlation
($R^2=0.38$).  (b) $k_f$ has excellent correlation with the
number of local minima ($R^2=0.85$), a property of the kinetic
landscape. The diamond represents the G\={o} model.}
\label{fig:thermo}
\end{figure}
\end{center}

Another thermodynamic property thought to be an important determinant
of folding rates is the collapse cooperativity $\sigma = 1-
T_f/T_\theta$ \cite{KlimovThirumalai95-PRL}, where $T_f$ is as defined
earlier, and $T_\theta$ the temperature when heat capacity $C(T)$
reaches its maximum.  Fig.~\ref{fig:thermo} shows that for the 26
sequences that fold to the same native structure in
Fig.~\ref{fig:seqMove}b, there is a weak correlation ($R^2=0.38$)
between collapse cooperativity and $\log k_f$.  Large variance in
observed folding rates exist for sequences of similar collapse
cooperativity.

\vspace*{-10mm}

The number of sequences that take a specific conformation as the
unique ground state is thought to be correlated with overall protein
stability and folding rates \cite{Melin}.  We calculated in addition
$k_f$ for a group of 79 singleton sequences with no sequence homologs
that fold to the same native conformations.  The distribution of
$k_f$s for the singleton sequences and the 26 sequences shown in
Fig.~\ref{fig:stability} demonstrate similarly large variation.  For
our model, designability is not an important determinant of the
folding rates.

The Inverse Participation Ratio $I$ is commonly used to characterize
the localization of eigenvectors.  It is defined as $I = \sum_k
v_k^2$, where $v_k$ is the $k$-th coefficient of the normalized
eigenvector.  The correlation between $I$ for the
equlibrium eigenvector and the folding rate for the 26 sequences is
 rather poor ($R^2=2\times10^{-3}$).

{\bf Kinematic determinants of folding landscape.}
Protein folding kinetics are intrinsically determined by physical movement of
molecules. Weak correlations of the folding rate with 
thermodynamic properties are not surprising.  Thermodynamic properties
of a sequence can be calculated if
its complete set of conformations are enumerated. Such properties are
not affected by the kinetic connections between conformations.  A
smooth energy landscape ensuring fast folding can be easily permuted
into a rugged landscape by assuming different transition rules between
conformations.  Both will have the same thermodynamic
properties, but the resulting folding rates for the same sequence will
be very different.  The energy landscape of folding is
dictated by the connection graph of states defined by the move
set. Characterizing such kinematic energy landscape is
therefore essential for studying protein folding dynamics.

Although the energy landscape contains 802,075 conformations, each is
connected by the move set to only a limited number ($\sim$ 30) of
conformations. We can identify states that are local minima, {\it
i.e.}, all states connected to which by moves have higher energy.  A simple
characterization of the kinematic energy landscape is then the
number count $n_{\min}$ of the local minima.  Fig.~\ref{fig:thermo}b
shows that an excellent correlation of $\log k_f$ and $n_{\min}$
($R^2= 0.85$) can be found for the 26 HP sequences that fold into the
same conformation.

Our conclusions are not sensitive to temperature $T$.  When $T$ is
raised from $0.20$ to $0.21$ (equivalent to raising $T$ from $300K$ to
$315K$), we found that the folding rate $k_f$ of the 26 sequences all
increases.  Although $k_f$ for slow folder increases more (by a factor
of 2.0 versus a factor of $1.4$ for fast folders), $k_f$ at $T=0.21$
is well-correlated with $k_f$ at $T=0.20$.  The correlation
coefficients of $\log k_f$ with the number of local minima, collapse
cooperativity (Fig.~\ref{fig:thermo}), and other thermodynamic
parameters are essentially unchanged.

{\bf Time evolution and basin analysis.}
Monitoring the exact time evolution of all conformations simultaneously until
reaching equilibrium during folding is a challenging task.
Mathematically, the model of master equation is equivalent to a Markov
process, where the population vector of conformations at time
$t+k\Delta t$ is given by $\bp(t+k\Delta t) = \bM^k \bp(t)$, where
$\bM = \bI + \bR \cdot \Delta t$, $\bI$ being the identity matrix.
However, the $k$-time step Markov matrix $\bM^k$ rapidly becomes a
dense matrix, and following the time evolution of folding by a
straightforward matrix multiplication of ${\cal O}(|\Omega|^3\log k)$ steps
becomes impossible for a large matrix of size $|\Omega|=802,075$ and
$k=10^6 - 10^{10}$.  The analytical solution of $\bp(t) = \sum_i
C_i \bn_i e^{-\lambda_i t}$ through diagonalization is also
impractical, as it is only possible to calculate a few eigenvectors
and eigenvalues for a large matrix.

We seek an accurate solution without the approximation of
macrostates.  Taking advantage of the sparsity of the rate matrix
$\bR$, we follow the approach of Sidje \cite{Sidje-expokit} and use the
analytical solution of matrix exponential: 
$
\bp(t) = e^{\bR t}\bp(0),
$
where $e^{\bR t}$ is defined by the Taylor expansion $e^{\bR t} = \bI
+ t \bR + \frac{t^2}{2} \bR^2 + \cdots + \frac{t^k}{k!}\bR^k+\cdots$.
This expansion itself is impractical, as it also involves
large matrix product of increasing density.  Plus, the entries in the
matrix terms may have alternating signs and hence cause numerical
instability.  A better approach is to expand $e^{\bR t} \bp (0)$ in the Krylov
subspace ${\mathcal{K}}_m$
defined as: 
\begin{equation}
{\mathcal{K}}_m(\bR t, \bp(0)) \equiv \mbox{Span}\{
\bp (0), 
\cdots, (\bR t )^{m-1} \bp(0)\}.
\end{equation}
Denoting $||\cdot||_2$ as the 2-norm of a vector or matrix, our
approximation then becomes $\bp (t) \approx ||\bp (0)||_2 \bV_{m+1}
e^{t \overline{\bH}_{m+1}} \be_1$, where $\be_1$ is the first unit
basis vector, $\bV_{m+1}$ is a $(m+1)\times (m+1)$ matrix formed by
the orthonormal basis of the Krylov subspace, and
$\overline{\bH}_{m+1}$ the upper Heisenberg matrix, both computed from
an Arnoldi algorithm. The error can be bounded by ${\cal
O}(e^{m-t||\bR||_2}(t||\bR||_2/m)^m)$.  We now only need to compute
explicitly $e^{\overline{\bH}_{m+1} t}$. Because $m$ is much smaller
than 802,075, this is a simpler problem.  A special form of the
Pad\'{e} rational of polynomials instead of Taylor
expansion is used for this \cite{Sidje-expokit}:
$e^{t\overline{\bH}_{m+1}} \approx N_{pp}(t \overline{\bH}_{m+1}
)/N_{pp} (-t \overline{\bH}_{m+1} )$, where $N_{pp}(t
\overline{\bH}_{m+1} ) =\sum_{k=0}^p c_k (t \overline{\bH}_{m+1} )^k$
and $c_k = c_{k-1}\cdot \frac{p+1-k}{(2p+1-k)k}$.  In our calculation,
we select  $m=30$.

\begin{wrapfigure}{r}{0.3\textwidth}
\centering
\includegraphics[width=0.3\textwidth]{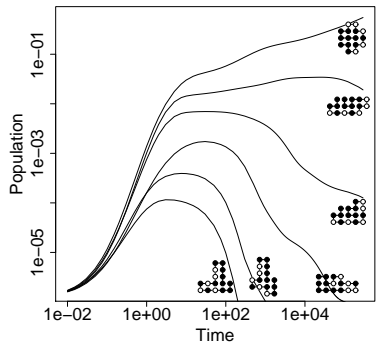}
\caption{\sf 
The time evolution of the native state and several local
minima states. The probability of occupation of native state
conformation (top) increases monotonically through a time span of
$10^{-2} - 10^{5}$, but local minima conformations go through transiently
accumulating intermediate states.}
\label{fig:evolution}
\vspace*{-3mm}
\end{wrapfigure}

Fig.~\ref{fig:evolution} shows an example of an HP sequence (sequence C in
Fig.~\ref{fig:seqMove}a) and the time evolution of its native
conformation and several local minima conformations. 
The time evolution of the native conformation shows an
initial fast phase upto $t \sim 50$  time units.
In principle, the local minima conformations can follow different
kinetic processes: Some could be transiently accumulating, and others
either monotopically accumulating or monotopically decreasing.  Based
on the computed trajectories of time evolution, we find that the
dynamic behavior of local minima conformations can be predicted from
{\it basin analysis\/} of the move-connected energy landscape.  We define the
size of the {\it basin\/} associated with each local minimum state $i$
computationally by artificially making every local minimum an
absorption state, {\it i.e.}, a sink of infinite depth, such that once
reached, no molecule can escape.  This is achieved by assigning
$r_{ij} = 0$ and $r_{ii}=1$ for each local minimum state $i$
\cite{ChangCieplak}.  $p'_{i}(t=\infty)$ therefore reflects the size
of the basin of the $i$-th local minimum.  We define the {\it
accumulation ratio\/} as
$
\varrho = \frac{p'_{i}(\infty)} {e^{-E_i/T}/\sum_j e^{-E_j/T}}.
$
If $\varrho>1$, state $i$ is most likely a transient accumulating state,
{\it i.e.}, the other conformations in its basin first rapidly flow to
state $i$ before transiting to conformations outside the basin.  If
$\varrho<1$, depending on its initial probability of occupancy and the
final Boltzmann factor, state $i$ may be either a monotonically
decaying or accumulating state.  We find that among the 493 local
minima states for this sequence, all except 3 are transiently
accumulating, indicating they are responsible for forming transient
state ensemble of various time scale.

To understand whether the formation of certain native contacts
facilitate folding, we examine the time evolution of each of the 8
native contacts ($a$--$h$) in Fig.~\ref{fig:seqMove}(b) for the 26
sequences.  We found that fast folders have larger amount native
contact $d$ ($R^2=0.74-0.81$ with $\log k_f$), and contact $c$ at the
transient time of$50-100$ (Fig.~\ref{fig:evolution}), indicating that
these contacts are critical for folding by restricting favorably the
conforamtional search space.  The formation of other native contacts
seem not to be directly related to folding rates.

To conclude, we studied protein folding dynamics using a model based
on detailed physical moves and exact solution of the master
equation. We found that folding rates vary enormously for sequences of
the same length, energy, energy gap, and even of the same
ground state conformation.  In contrast to the thermodynamic
parameters which are weak predictors of folding rates, properties of
the kinematic landscape defined by the physical moves provide
excellent correlation with folding rates.  With the computation of
time evolution of individual conformation from the first move to
half-time of equilibrium, we show that many transiently accumulating
intermediate states can be identified by basin analysis.

\addtolength{\textheight}{-3cm}   

\begin{acknowledgements}
We thank Drs.\ Ken Dill, Bosco Ho, Xiaofan Li, Banu Ozkan, Dev
Thirumalai, Jin Wang, and Weitao Yang for helpful discussions.  This
work is supported by NSF DBI0133856, NIH GM68958, and Whitaker
TF-04-0023.
\end{acknowledgements}
\vspace*{5mm}
$^*$ Corresponding author.  Email: {\tt jliang@uic.edu}

\end{document}